\documentclass[openacc]{rstransa}

\usepackage{graphicx}
\usepackage{bbold}
\usepackage{epstopdf, epsfig}
\usepackage{amsmath,amssymb,mathtools}
\usepackage{braket}
\usepackage[english]{babel}
\usepackage{ulem}
\usepackage{cancel}

\newcommand{\beq}{\begin{equation}}
\newcommand{\eeq}{\end{equation}}
\newcommand{\bu}{{\bf u}}

\def\bk{{\bf k}}

\def\cP{ \mathcal{P} }

\def\cE{ \mathcal{E} }

\def\NEW#1{\textcolor{black}{#1}}

\usepackage{textgreek}

\begin{document}

\title{Geometric microcanonical theory of two-dimensional Truncated Euler flows}

\author{
A. van Kan$^{1}$, A. Alexakis$^{1}$ and M. Brachet$^{1}$}

\address{$^{1}$ Laboratoire de Physique de l'Ecole normale sup\'erieure, ENS, Universit\'e PSL, CNRS, Sorbonne Universit\'e, Universit\'e de Paris, F-75005 Paris, France}

\subject{Fluid Dynamics, Statistical Mechanics}

\keywords{microcanonical ensemble, two-dimensional flow, truncated Euler equations, turbulence, reversals}

\corres{Adrian van Kan\\
\email{adrian.van.kan@phys.ens.fr}}

\begin{abstract}
This paper presents a geometric microcanonical ensemble perspective on two-dimensional Truncated Euler flows, which contain a finite number of (Fourier) modes and conserve energy and enstrophy. We explicitly perform phase space volume integrals over shells of constant energy and enstrophy. Two applications are considered. In a first part, we determine the average energy spectrum for highly condensed flow configurations and show that the result is consistent with Kraichnan's canonical ensemble description, despite the fact that no thermodynamic limit is invoked. In a second part, we compute the probability density for the largest-scale mode of a free-slip flow in a square, \NEW{which displays reversals}. We test the results against numerical simulations of a minimal model and find excellent agreement with the microcanonical theory, \NEW{unlike the canonical theory, which} fails to describe the bimodal statistics.
This  article is part of  the  theme  issue  ``Mathematical problems in physical fluid dynamics".
\end{abstract}


\begin{fmtext}
\section{Introduction}
Turbulent flows involve a large number of degrees of freedom, spanning many spatial and temporal scales. Similarly, in a gas at equilibrium, there is a large number of degrees of freedom corresponding to all the gas molecules. In the latter case, it is well known that equilibrium statistical mechanics provides a description of drastically reduced complexity. Turbulent flows are, however, non-equilibrium phenomena \cite{goldenfeld2017turbulence}, since they involve finite fluxes of energy and other invariants across scales due to nonlinear interactions. For instance, energy is transferred from large to small scales in homogeneous and isotropic three-dimensional turbulence \cite{frisch1995turbulence} whereas in two dimensions it flows from small to large scales \cite{bofetta2012twodimensional}. At first sight, this makes 
the two cases starkly different.
\end{fmtext}


\maketitle

\noindent However, despite turbulence being an out-of-equilibrium phenomenon overall,  equilibrium theory does remain relevant under certain circumstances. In three dimensions, this has been claimed to be the case at scales larger than the injection scale. At these scales, the energy flux is zero and the system can possibly be modeled using equilibrium dynamics \cite{forster1977large,	dallas2015statistical, cameron2017effect, alexakis2018thermal}. In addition, understanding equilibrium dynamics is important for systems  that  display  a  transition  from a forward  to  an  inverse  cascade  \cite{alexakis2018cascades, sahoo2017discontinuous, benavides2017critical, deusebio2014dimensional, seshasayanan2014edge,van2019condensates, van2019critical,vankan2021energy}; in these systems the large scales transition from an equilibrium state to an out-of-equilibrium state. Another instance of equilibrium properties in three-dimensional turbulence is the so-called bottleneck, which manifests at the smallest scales of the inertial range (the range of scales below the forcing scale and above the dissipation scales), where the power-law spectrum becomes less steep \cite{donzis2010bottleneck, falkovich1994bottleneck, martinez1997energy, lohse1995bottleneck}. The bottleneck was interpreted as \textit{incomplete thermalisation} in \cite{frisch2008hyperviscosity}, where it was argued that the scales involved in the bottleneck are asymptotically at equilibrium for hyper-viscous flows as the order of the hyper-viscosity goes to infinity. This prediction was recently shown to be consistent with numerical evidence \cite{agrawal2020turbulent}. 

Arguably the most successful application of equilibrium statistical mechanics to turbulence has been the case of two-dimensional (2-D) flows in finite domains, \NEW{where energy accumulates in the mode(s) associated with the largest available spatial scale, forming a so-called condensate} \cite{kraichnan1967inertial,kraichnan1980two,robert1991statistical, naso2010statistical, bouchet2012statistical,shukla2016statistical}. An important property of such 2-D turbulent flows is that, in contrast with three dimensions, the energy dissipation vanishes when the viscosity tends to zero.  \textcolor{black}{Thus energy fluxes through the system also vanish in that limit \cite{tabeling2002two}. In addition to the energy, 2-D Euler flow also conserves the integral of the square of vorticity, known as enstrophy. Whether the dissipation of enstrophy vanishes in the zero-viscosity limit of the 2-D Navier-Stokes equations is known to depend on the choice of forcing mechanism for forced 2-D turbulence \cite{eyink2001dissipation,alexakis2006energy}. For instance, monochromatic and constant-injection-rate forcing leads to vanishing enstrophy dissipation as viscosity goes to zero. For decaying 2-D turbulence, the answer depends on the initial conditions having finite enstrophy or not \cite{tran2005enstrophy,tran2006vanishing}. 
In either case,  at scales larger than the forcing scale both energy and enstrophy fluxes vanish at steady state
\cite{kraichnan1967inertial}.
Thus these scales may be considered to be in equilibrium. 
} 

Two main approaches from statistical physics can be considered to describe such flows, which will be described in more detail below. Firstly the microcanonical ensemble, which applies to closed systems, and secondly the canonical and grand canonical ensembles, which apply to open systems subject to fluctuations of energy and other quantities (typically particle number) around a mean value. The first attempt in this direction was undertaken by Onsager in 1949 \cite{onsager1949statistical}, who formulated a microcanonical description of idealised (singular) point-vortex flow to explain the self-organisation of 2-D turbulence (see \cite{eyink2006onsager} for a review of Onsager's contributions to turbulence). Since Onsager's initial contribution, the statistical mechanics of singular point vortices has continued to attract a great deal of attention \cite{joyce1973negative, lundgren1977statistical, caglioti1992special, eyink1993negative, kiessling1997micro, chavanis2012kinetic, esler2013statistical, dritschel2015ergodicity, esler2017equilibrium, eyink2006onsager}. 
%
A generalization of the point-vortex statistical description
was proposed by the celebrated Robert-Sommeria-Miller (RSM) theory proposed in the early 1990s \cite{robert1991maximum, miller1990statistical,robert1991statistical,miller1992statistical, eyink2006onsager,chavanis2009dynamical}. The full 2-D Euler equations conserve vorticity for every fluid parcel. Hence the integral of any power of vorticity is conserved, not only the enstrophy. This implies an infinite family of conserved quantities (known as Casimir invariants), which was taken into account. A detailed description of RSM theory and its further developments can be found in \cite{bouchet2012statistical}, a concise introduction is also given in \cite{eyink2006onsager}. The basic object of the theory is a local ``microscopic'' distribution function $n(\sigma,\mathbf{r})$, the probability density associated with vorticity $\omega(\mathbf{r})$ lying between $\sigma$ and $\sigma+d\sigma$ at the space point $\mathbf{r}$. The idea is that after evolving for a long time, the vorticity field develops very fine scales so that a small neighborhood of the point $\mathbf{r}$ will contain many values of the vorticity with levels distributed according to $n(\sigma,\mathbf{r})$. From this distribution, a maximum principle for a generalised entropy leads to a mean-field equation for the ``macroscopic'' stream function, whose solution yields the equilibrium flow configuration. Specifically, RSM theory has been successfully applied to Jupiter's Great Red Spot \cite{bouchet2000emergence}, ocean rings and currents \cite{venaille2011oceanic} and zonal flows \cite{bouchet2016zonal}.


Here we follow an alternative equilibrium statistical description of turbulence, which can be obtained by considering
the equilibrium state of the truncated (incompressible) Euler equations (TEE). The TEE retain only a finite number of Fourier modes
\cite{lee1952some,hopf1952statistical,kraichnan1973helical,orszag1974lectures}. \textcolor{black}{When the Euler equation is studied numerically, for instance with periodic boundary conditions, these are precisely the equations which \NEW{pseudo-spectral numerical codes} solve.}
In 1952, Lee \cite{lee1952some} investigated this system and showed that the TEE satisfy Liouville's theorem \textcolor{black}{of conservation of phase space volume}.  In three dimensions, assuming ergodicity, Lee \cite{lee1952some}  predicted that at equilibrium this system  will be such that every state $\mathbf{u}$ of a given energy $\mathcal{E}$ is equally probable. This leads to the prediction that the energy spectrum ${E}(\mathbf{\bk})$ (defined as the mean energy of the \NEW{wave vector} $\bk$) is given by $E(\bk) =\mathcal{E}/N$, where $N$ is the total number of \NEW{wave vectors}. This is equivalent to the microcanonical ensemble  in  statistical physics, \textcolor{black}{which has been extensively studied for small systems \cite{gross1997microcanonical,gross2001microcanonical}}, and it here amounts  to  equipartition  of  energy  among  all  the degrees of freedom (i.e. among all Fourier amplitudes).
Two decades later, Kraichnan \cite{kraichnan1973helical} considered the TEE, for which he proposed a different approach, 
by considering that the complex amplitudes of the Fourier modes followed a canonical distribution that was controlled by the mean values of the invariants of the system: energy and helicity. Kraichnan's approach corresponds to a grand canonical ensemble, as total energy and helicity 
are allowed to fluctuate around  a mean value. The grand canonical approach allowed Kraichnan to generalise Lee's result to a modified energy spectrum in the presence of helicity. A review of these results can be found in  \cite{orszag1974lectures}. 
We note that a microcanonical statistical description of  \textcolor{black}{finite-dimensional} 3-D TEE taking into account both the energy and the helicity constraint, has not been achieved. This is because, as we will see for the 2-D case, the presence of an additional invariant significantly complicates the integrals involved.

In two dimensions the TEE and the grand canonical ensemble statistics were investigated again by Kraichnan \cite{kraichnan1975statistical}. 
The 2-D TEE  can be written in terms of the stream function $\psi(\mathbf{r})$ at position $\mathbf{r}$ (related to velocity via $\mathbf{u}=\hat{e}_3\times\nabla\psi$), 
\begin{equation}
\partial_t \omega + \mathbb{P}_K J(\psi,\omega) = 0,  \label{eq:tee}
\end{equation}
\noindent where $\omega=\nabla^2 \psi$ is vorticity, $J(f,g)=(\partial_xf)( \partial_y g) - (\partial_x g)( \partial_y f)$ is the Jacobian operator, $x,y$ are the space coordinates, and  $\mathbb{P}_K$ is a projection operator that sets equal to zero all Fourier modes except those that belong to a particular set $K$.
The TEE possess exactly two invariants, namely
\begin{equation}
    \text{ energy } \mathcal{E} =
\frac{1}{2}\int|\mathbf{u}|^2 d^2x, 
\hspace{1cm} \text{ and enstrophy } 
{\Omega} = 
\frac{1}{2}\int|\nabla\times \mathbf{u}|^2 d^2x.  
\label{eq:def_energy_enstrophy}
\end{equation}
In Fourier space, energy and enstrophy are distributed over the different modes. This is quantified by the 2-D energy spectrum,
which, in terms of the Fourier transform $\hat{\psi}(\mathbf{k})$ of $\psi(\mathbf{r})$, reads 
\begin{equation}
E(\mathbf{k}) = \frac{1}{2} k^2|\hat{\psi}(\mathbf{k})|^2. 
\end{equation}
Note that $E(\mathbf{k})$ is the energy contained in the single mode with wave vector $\mathbf{k}$, and is not summed over the wave number shell of radius $|\mathbf{k}|$, by contrast with the commonly used isotropic energy spectrum. At late times the solution of the TEE reaches a statistically steady state whose properties are fully determined by $\mathcal{E}$ and $\Omega$. 
 Kraichnan's \cite{kraichnan1975statistical} grand canonical ensemble assumes again that the Fourier amplitudes follow a canonical distribution:
\begin{equation}
P(\mathbf{u}) = \mathcal{Z}^{-1}\exp(-\alpha \mathcal{E} - \beta \Omega),
    \label{eq:kraichnan_canonical_pdf}
\end{equation}
where $\mathcal{Z}$ is a normalisation constant, $P(\mathbf{u})$ is the probability density function (PDF) associated with the system having the velocity field $\mathbf{u}$. The constants $\alpha,\beta$ are Lagrange multipliers, analogous to inverse temperature and inverse chemical potential in a gas at equilibrium. It implies the average energy spectrum $E(\mathbf{k}) =(\alpha+\beta k^2)^{-1}$. Note, that (\ref{eq:kraichnan_canonical_pdf}) is not exact for the TEE, since it allows for fluctuations of energy and enstrophy, which are invariants of the TEE. 

The alternative is to assume only ergodicity and use the microcanonical description of Lee \cite{lee1952some}. 
This amounts to attributing uniform probability in \NEW{the subset of} phase space that satisfies the energy constraint $\cE=\cE_0$ and enstrophy constraint $\Omega=\Omega_0$ where $\cE_0$ and $\Omega_0$ is the initial energy and enstrophy of the system.
Explicitly, the PDF is given by:
\begin{equation}
    P(\mathbf{u}) = \mathcal{Z}^{-1} 
    \delta(\mathcal{E}-\mathcal{E}_0) \delta(\Omega-\Omega_0),
    \label{eq:def_micro_pdf}
\end{equation}
with normalisation $\mathcal{Z}$ (different from that in (\ref{eq:kraichnan_canonical_pdf}))
and $\mathcal{E}$ and $\Omega$ defined in (\ref{eq:def_energy_enstrophy}). In geometrical terms, this distribution \NEW{in phase space} is non-zero only at the intersection between  
the manifold determined by the energy constraint  
and the manifold
determined by the enstrophy constraint in the $N$-dimensional phase space. Not surprisingly, it is non-trivial to obtain 
analytical results using the microcanonical ensemble, as the integrals involved in the computation of any mean quantity have
to be performed over a high-dimensional (co-dimension 2), complicated submanifold in phase space. \textcolor{black}{Both the microcanonical and canonical ensembles correspond to invariant measures, in the sense that they are stationary solutions of Liouville's equation for the probability density, because they both depend on invariants only \cite{orszag1974lectures}.}

In general, working in the canonical ensemble greatly simplifies computations. While it is found in many cases that the canonical results asymptotically agree with the microcanonical ones as the number of degrees freedom tends to infinity (the {\it thermodynamic limit}), there are also examples of ensemble inequivalence in this limit, in particular in systems with long-range interactions, \cite{lewis1994equivalence,ellis2000large,ellis2002nonequivalent,venaille2009statistical,venaille2011solvable,bouchet2012statistical}. Moreover, for systems in which the energy is concentrated in only a small number of modes, as is the case in large-scale condensates, there is a priori no reason to expect the two statistical ensembles to yield the same result. In this case, for exactly conservative systems such as the TEE, the micro-canonical ensemble is the more appropriate choice, since it respects the conservation laws and only assumes the dynamics to be ergodic. Therefore, despite the technical difficulty it entails, the study of the microcanonical ensemble is highly relevant to the TEE. 

In this work, we propose a novel approach to the microcanonical statistical mechanics of TEE flows. We explicitly compute the intersection volume and deduce different statistical quantities based on the microcanonical distribution (\ref{eq:def_micro_pdf}) for two examples. First, we consider a condensate flow and compute the microcanonical average energy spectrum. Second, we extend the work of \cite{shukla2016statistical} to show that the statistics of reversals of the largest-scale velocity in a simple free-slip flow in a square domain are correctly predicted by an explicitly geometrical microcanonical calculation.  

\section{ Energy spectrum of condensate flows}             
\label{sec:cond}
\subsection{Microcanonical calculation}
\label{sec:cond_mic}
In this section we calculate the energy distribution among modes.
Consider a 2-D flow with boundary conditions leading to a discrete set of Fourier modes, e.g. in a periodic domain,
\begin{equation}
    \psi(\mathbf{r}) = \sum_{\mathbf{k}\in K} \hat{\psi}(\mathbf{k}) e^{i\mathbf{k}\cdot \mathbf{r}}
\end{equation}
 with complex amplitudes $\hat{\psi}(\mathbf{k})$ satisfiying the condition $\hat{\psi}(-\bk)=\hat{\psi}^*(\bk)$, required for $\psi$ to be real, the summation being over the set $K= \left\lbrace 2\pi\left(\frac{ n}{L_x},\frac{ m}{L_y} \right) \left| (n,m)\in \mathbb{Z}^2\right. \right\rbrace \cap \left\lbrace \mathbf{k}\in \mathbb{R}^2\left|0<|\mathbf{k}|\leq k_{max} \right.\right\rbrace$ for a domain size $L_x\times L_y$, or to a discrete set of sine modes, e.g. for a $[0,\pi]^2$ free-slip domain,
\begin{equation}
\psi(\mathbf{r}) = \sum_{\mathbf{k}=(n,m)\in K} \hat{\psi}(\mathbf{k}) \sin(m x)\sin(ny),   
\end{equation}
with real amplitudes $\hat{\psi}(\mathbf{k})$ depending on $\mathbf{k}$ in $K=\left \lbrace \mathbf{k}=(m,n) \left| 0<|\mathbf{k}|<k_{max}; m,n\in \mathbb{N}_+ \right. \right \rbrace$. In the following, we always denote by $N$ the number of elements in the set $K$, independently of whether the amplitudes $\hat{\psi}(\mathbf{k})$ are real or complex. If the amplitudes are complex, the real and imaginary parts of $\hat{\psi}$ are separate degrees of freedom, but only for half the wave vectors. 
In either case (real or complex amplitudes) the number of degrees of freedom is equal to the number of wave vectors $N$.
We label the degrees of freedom by an index $i=1,\dots,N$, and denote the associated wave vector by $\mathbf{k}_i$, with wavenumber $k_i = |\mathbf{k}_i|$. 
\textcolor{black}{In order to obtain a real-valued phase space whose components are indexed in such a way that the corresponding wavenumber is a non-decreasing function of the index,} we introduce the following new variables: if $\hat{\psi}(\mathbf{k})$ is real, then $r_i\coloneqq\hat{\psi}(\mathbf{k}_i) k_i /\sqrt{2}$, and the index $i$ covers all wave vectors. If $\hat{\psi}$ is complex, then 
$r_{i} \coloneqq \mathrm{Re}\lbrace\hat{\psi}(\mathbf{k}_i)\rbrace k_i$ if $i$ is even,
$r_{i} \coloneqq \mathrm{Im}\lbrace\hat{\psi}(\mathbf{k}_i)\rbrace k_i$ if $i$ is odd and $i$ covers half the wave vectors so that $\bk_i$ and $-\bk_i$ together
cover all wave vectors. The labeling is such that the $k_i$ are ordered so that $k_i \le k_{i+1}$ and let also $k_1=k_2=\dots=k_{_M}< k_{_{M+1}}$ be the first $M$ equal smallest wavenumbers.
For instance, in a periodic square spatial domain $[0,2\pi]^2$, $M=4$ with $k_1=\dots=k_4=1$, corresponding to the real and imaginary parts of $\mathbf{k}=(1,0),(0,1),(-1,0),(0,-1)$, taking into account that $\psi$ is real. For free-slip boundary conditions in a $[0,\pi]^2$ domain, one finds $M=1$ with $k_1=\sqrt{2}$. 

Geometrically, with this notation, constant-energy trajectories in phase space satisfy $\sum r_i^2=\cE$,  i.e. they live on the surface of an N-dimensional sphere of radius $\sqrt{\cE}$. Constant-enstrophy trajectories follow $\sum k_i^2 r_i^2=\Omega$ and thus live on the surface of an N-dimensional ellipsoid with
the longest ellipse semi-axis is $\Omega^{1/2}/k_1$, the shortest semi-axis is $\Omega^{1/2}/k_{_N}$. The two hyper-surfaces intersect when $\cE k_1^2 \le \Omega \le \cE k_{_N}^2 $. Phase space trajectories of the TEE that conserve both energy and enstrophy thus live on this intersection of the two hyper-surfaces. \textcolor{black}{Note that this is the $N$-dimensional analogue of the energy and angular momentum conservation for a freely spinning top (see \S 37 of \cite{landau1976mechanics}).}

Our goal is to calculate the temporal mean energy spectrum $E(\bk_i)=\langle r_i^2\rangle$ for a flow with initial energy $\cE$ and enstrophy $\Omega$.  The assumption of ergodicity allows us to replace the temporal mean by an average over phase space volume, thus
\beq
\langle r_i^2\rangle =\frac{1}{\mathcal{Z}} \int r_i^{2} \delta\left( \sum_{j=1}^N r_j^2- \cE \right) \delta\left(\sum_{j=1}^N r_j^2k_j^2 - \Omega \right) \prod_j dr_j,
\label{eq:r2}
\eeq
where
\beq
\mathcal{Z} = \int \delta\left( \sum_{j=1}^N r_j^2- \cE \right) \delta\left(\sum_{j=1}^N r_j^2k_j^2 - \Omega \right) \prod_j dr_j.
\label{eq:z}
\eeq
\noindent 
In particular, we are interested in the limiting case where
\beq  \Omega = \cE k_1^2 (1+ \epsilon^2), \quad  \mathrm{with} \quad \epsilon \ll 1, \label{eq:EnOm1}\eeq
such that almost all energy is concentrated in the small-$k$ modes. 
\NEW{This case is closely related to the situation met in forced 2-D turbulence, where the inverse cascade leads to
a high condensation of energy at the smallest wavenumbers, displaying quasi-equilibrium statistics. Also, in this case,
because energy is concentrated in a few modes, the thermodynamic limit $N\to\infty$ could fail.
A priori we cannot tell if the two limits $\epsilon\to 0$ and $N\to\infty$ commute. } 

\begin{figure*}                                                      
\includegraphics[width=0.45\textwidth]{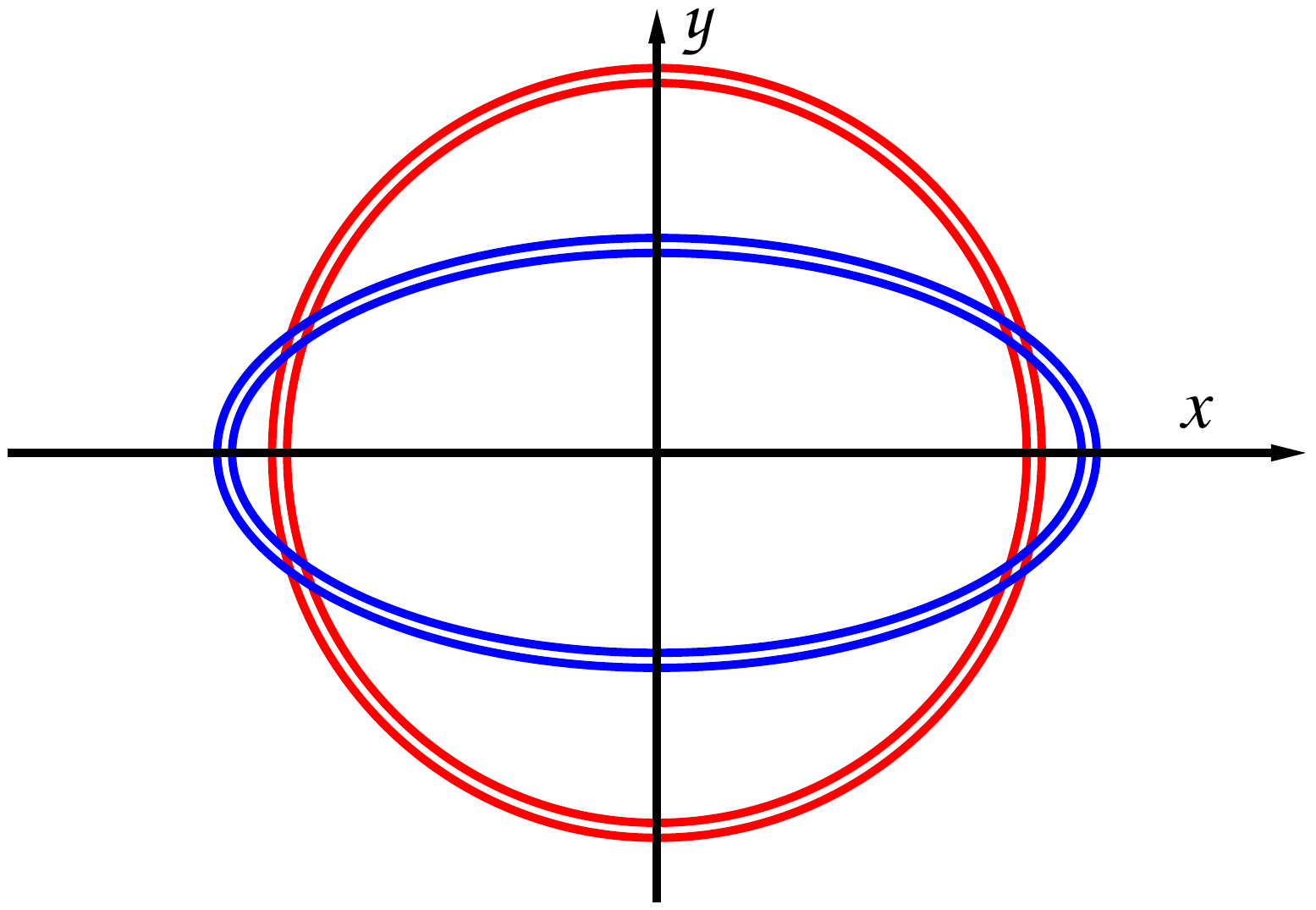}                  
\includegraphics[width=0.40\textwidth]{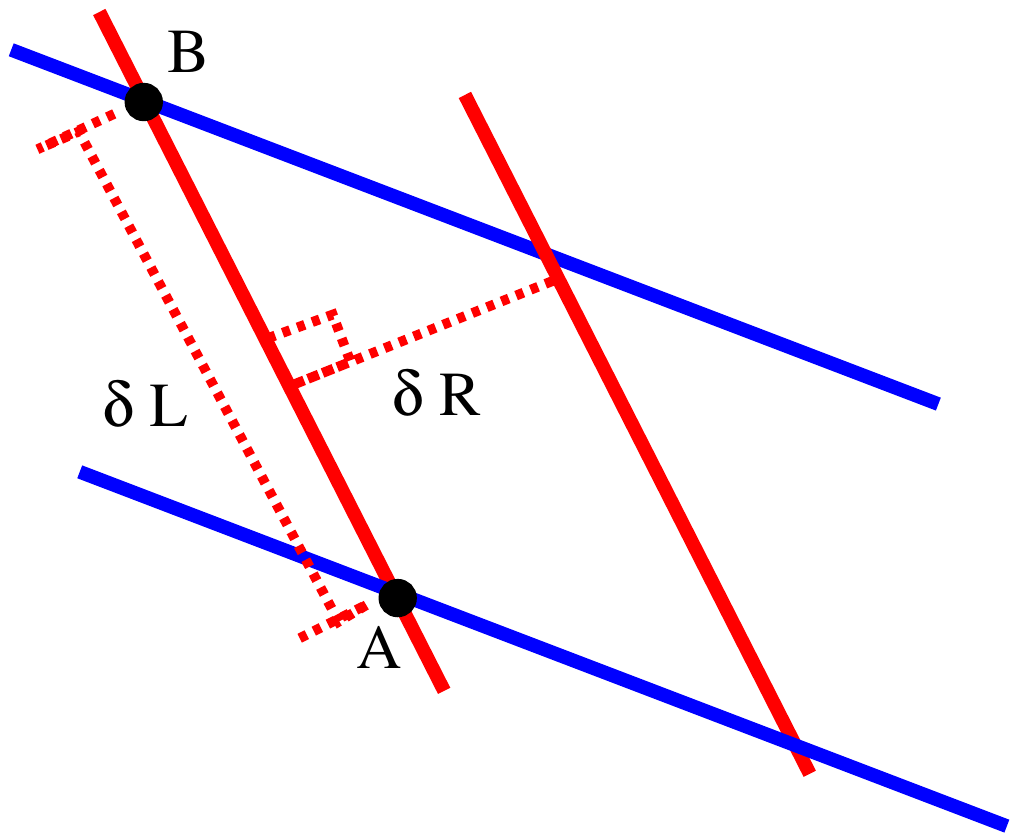}                    
\caption{Left: 
A cross-section of the geometry studied here. The greatest semi-axis of the ellipsoid is slightly larger than the sphere's radius, such that the largest-scale modes $k_1,\dots,k_M$ are $O(1)$, while $r_{M+1},\dots,r_N$ are $O(\epsilon)$.
Right: Zoom on the intersection area. \label{fig:circles_ellipses}}  
\end{figure*}

\NEW{In geometrical terms
$\epsilon\ll1$} means that the largest ellipse semi-axis, $\Omega^{1/2}/k_1$, is slightly larger than the sphere radius, $\cE^{1/2}$, as sketched in 
the left panel of figure \ref{fig:circles_ellipses}. The delta functions restrict the integrals to values of $r_1,r_2,\dots,r_{_M} \in [-\cE^{1/2},\cE^{1/2}]$, to be of order one, while $r_{_{M+1}},r_{_{M+2}},\dots,r_{_N}$ are of order $\epsilon$. \textcolor{black}{We define $x^2 \coloneqq \sum_{i=1}^{M} r_i^2$ as the energy in the largest scale and $y^2 \coloneqq \sum_{i=M+1}^N r_i^2$ as the energy in the remaining scales}. The equations then become
\beq  x^2     + y^2                = \cE \quad \mathrm{and } \quad 
k_1^2 x^2     + q_{_{M+1}}^2(\Phi) y^2          = \cE k_1^2(1+\epsilon^2), \label{eq:sphell}\eeq
where
\begin{eqnarray}
q_{_{M+1}}^2(\Phi)   &=& \left. \left(\sum_{n=M+1}^N k_n^2 r_n^2\right) \middle/ \left(\sum_{n=M+1}^N  r_n^2\right)\right. 
 \label{eq:def_q}
\end{eqnarray}

%
%
\textcolor{black}{Note that $q^2_{M+1}(\Phi)$ depends on the values $r_{M+1},\dots,r_N$, but not on $x$ and $y$. Using spherical coordinates in the subspace $(r_{M+1},\dots,r_N)$, it can be expressed in terms of a set of angles $\Phi$. Similarly, for the subspace $(r_1,\dots,r_M)$, we introduce another set of spherical coordinates, with a set of angles denoted by $\Theta$. The transformation to the two spherical coordinate systems is given in appendix \ref{sec:appA}.} The values of $x^2$ and $y^2$ that satisfy (\ref{eq:sphell}) can be then be expressed in terms of $q_{_{M+1}}(\Phi)$ as
\beq
x^2 = \cE -\frac{k_1^2\epsilon^2 \cE}{q_{_{M+1}}^2-k_1^2} ,\qquad y^2 = \frac{k_1^2\epsilon^2 \cE}{q_{_{M+1}}^2-k_1^2} .
\label{eq:xyq}
\eeq

To compute the integrals in (\ref{eq:r2}), ($\ref{eq:z}$), we fix the angle coordinates $\Theta,\Phi$ (\textcolor{black}{and thus the $q_{M+1}(\Phi)$}), and
consider the volume with energy in the range $[\cE,\cE+d \cE]$ and enstrophy in $[\Omega,\Omega+d\Omega]$, with $d \cE$, $d\Omega$ infinitesimal.
Then, in the $x,y$ plane the N-spherical shell and the N-ellipsoidal shell intersect 
forming a parallelogram, shown in figure \ref{fig:circles_ellipses}, of height 
$\delta R = d \cE\Big/\left(2\sqrt{\cE}\right)$ 
 and base length $\delta L$ defined as the distance between the points $(x_{_A},y_{_A})$ and $(x_{_B},y_{_B})$ given by
 the intersection of the curves $x_{_A}^2+y_{_A}^2=\cE$ and $k_1^2 x_{_A}^2+q_{_{M+1}}^2y_{_A}^2=\Omega$,
 and $x_{_B}^2+y_{_B}^2=\cE$ and $k_1^2 x_{_B}^2+q_{_{M+1}}^2y_{_B}^2=\Omega+d \Omega$, respectively.
A straightforward Euclidean calculation gives, to first order in  $\epsilon$,
\begin{eqnarray}
\delta L^2 &=& (y_{_B}-y_{_A})^2 + (x_{_B}-x_{_A})^2 \\
           &=     & \left( \frac{d\Omega}{2(q_{M+1}^2-k_1^2)} \right)^2 \left( \frac{1}{x_A^2}+\frac{1}{y_A^2}\right) \\
           &\simeq&        \frac{d \Omega^2}{4(q_{_{M+1}}^2-k_1^2)^2y_A^2}   , 
\end{eqnarray}
The area of the parallelogram is thus, to leading order,
\begin{eqnarray}
\delta A = \delta R \delta L 
&\simeq & \frac{d \cE d \Omega}{4 \cE^{1/2} (q_{_{M+1}}^2-k_1^2)|y_A|} 
        = \frac{d \cE d \Omega}{4 \epsilon \cE k_1 \sqrt{q_{_{M+1}}^2-k_1^2} } .
\end{eqnarray} 
\textcolor{black}{The remaining part of the calculation amounts to integrating this infinitesimally small area element over all remaining degrees of freedom (i.e. the angles $\Phi,\Theta$). This rather lengthy, but straightforward calculation is done in appendix \ref{sec:appA}, and gives that for all $i>M$,}
\beq
E(\bk_i)=
\langle r_i^2 \rangle =  \frac{\epsilon^2 \cE k_1^2 }{(N-M)(k_i^2-k_1^2)}.
\label{eq:In_i_gt_M}
\eeq
Conservation of energy thus yields, at leading order,
\beq
E(\bk_i)= \langle {r_i}^2 \rangle =  \frac{1}{M}\left( \cE - \sum_{j=M+1}^{N} \langle r_j^2 \rangle \right)  = \cE/M + O(\epsilon^2). \label{eq:In_result_lt_M}
\eeq

\subsection{Comparison with Kraichnan's canonical ensemble prediction} 
From Kraichnan's canonical ensemble  probability density (\ref{eq:kraichnan_canonical_pdf}), one can compute the canonically averaged \NEW{2-D} energy spectrum \NEW{(energy of the single mode with wave vector $\mathbf{k}_i$)}
$\langle r_i^2 \rangle_c$. One finds
\begin{equation}
    E(\bk_i)  = \frac{1}{2(\alpha+\beta k_i^2)},
\end{equation}
with $\alpha,\beta$ determined by $\cE = \frac{1}{2} \sum_i E(\bk_i)$, and $\Omega = \frac{1}{2} \sum_i k_i^2 E(\bk_i)$. For highly condensed flows, where $E(\bk_i)\gg E(\bk_j)$ for any $i\leq M$, $j\geq M+1$, one requires $\alpha/\beta = -k_1^2 (1-\delta^2),$ $\delta \ll 1$. Hence
\begin{eqnarray}
\cE =& \frac{M}{2\beta\delta^2 k_1^2} + \frac{1}{2\beta}\sum_{i=M+1}^N  \frac{1}{k_i^2-k_1^2(1-\delta^2)} &= \frac{M}{2\beta\delta^2 k_1^2} + \text{ higher-order terms},\, \\ \, \Omega =& \frac{M}{2\beta\delta^2}+\frac{1}{2\beta}\sum_{i=M+1}^N  \frac{k_i^2}{k_i^2-k_1^2(1-\delta^2)}&= \frac{M}{2\beta\delta^2} +  \text{ higher-order terms},
\end{eqnarray}
where again $M$ is the number of modes with $|\mathbf{k}|=k_1$. These expressions imply that $\Omega=k_1^2 \cE + O(1)$, and $\beta^{-1} =\frac{2\cE k_1^2\delta^2}{M}$. Furthermore, using the definition of $\epsilon$ in eq. (\ref{eq:EnOm1}), we find that, at leading order $\epsilon^2 \sim \frac{\delta^2(N-M)}{M} $. This gives at leading order
\begin{equation}
    \langle r_i^2 \rangle_c = \begin{cases} \frac{\cE}{M} \qquad &: 1\leq i \leq M \\ \frac{\cE \epsilon^2 k_1^2}{(N-M)(k_i^2-k_1^2)}  &: M+1\leq i\leq N, \end{cases}
\end{equation}
which is identical to the microcanonical results, although the latter involved no thermodynamic large-$N$ limit, but only a small-$\epsilon$ limit. 

\NEW{The agreement of the two calculations indicates that the two limits $\epsilon\to0$ and $N\to\infty$ commute in this case.}
\NEW{The microcanonical result provides an added value, since it is valid for any $N$, even in the absence of the thermodynamic limit, under the hypothesis of ergodicity}. In the condensate state examined here, where most of the energy is concentrated in few modes, there is no guarantee that the grand canonical result applies. In fact, in the example presented in the next section, we show that the microcanonical and grand canonical ensembles give different results.

\section{Reversals in free-slip flow in the square domain}   
\subsection{Microcanonical calculation}
\begin{figure}
     \centering
     \includegraphics[width=0.325\textwidth]{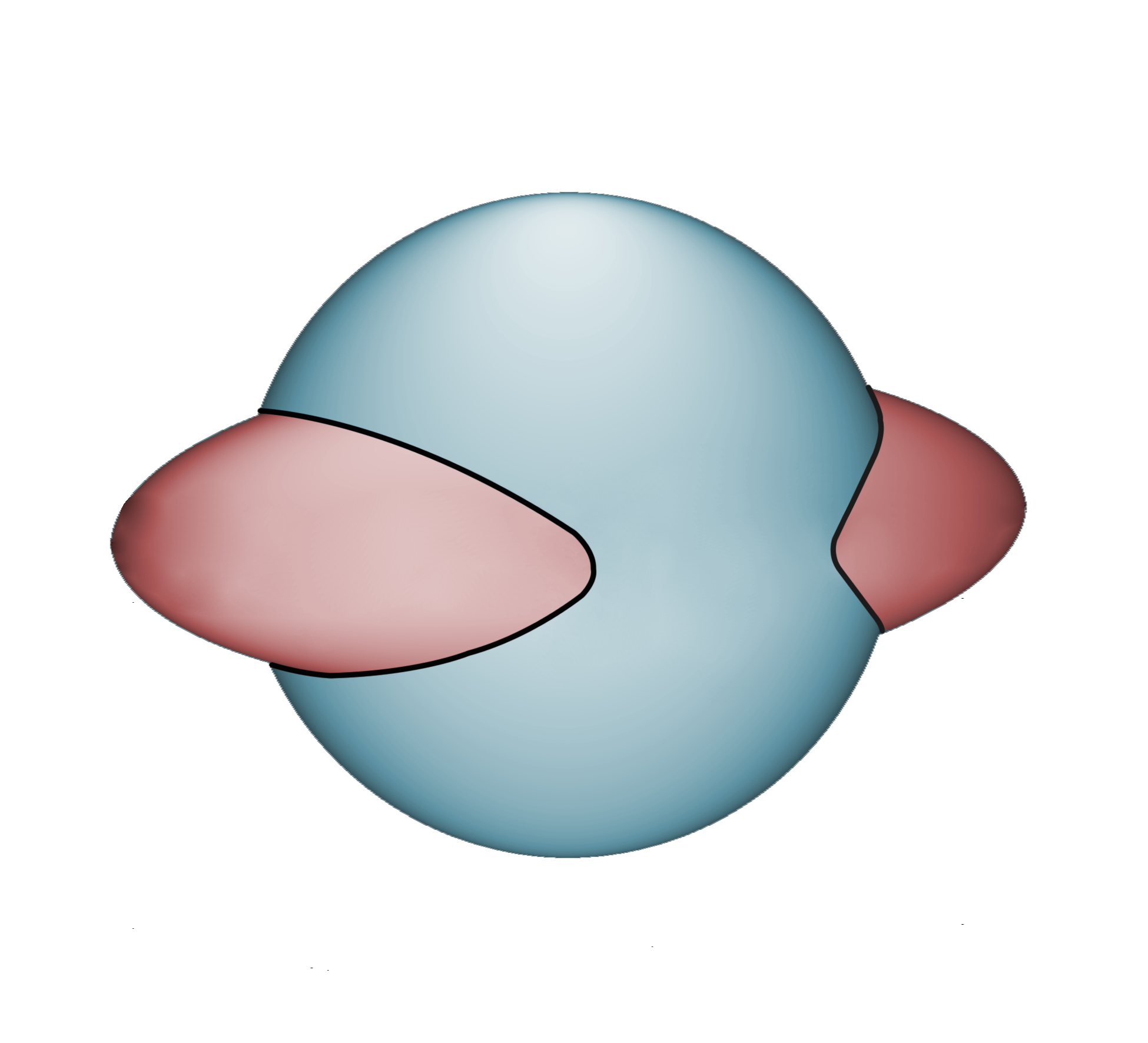}
     \includegraphics[width=0.325\textwidth]{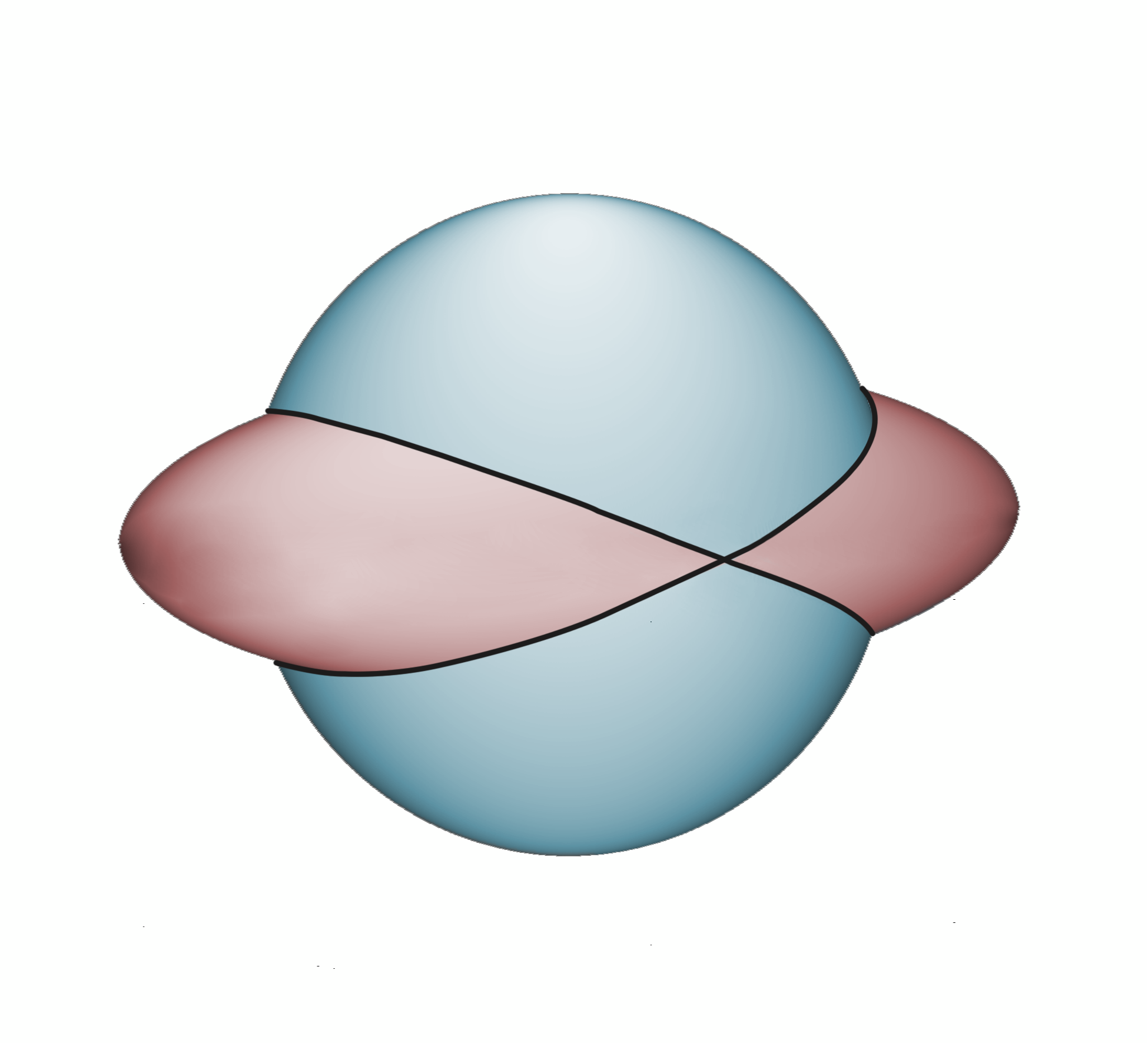}
     \includegraphics[width=0.325\textwidth]{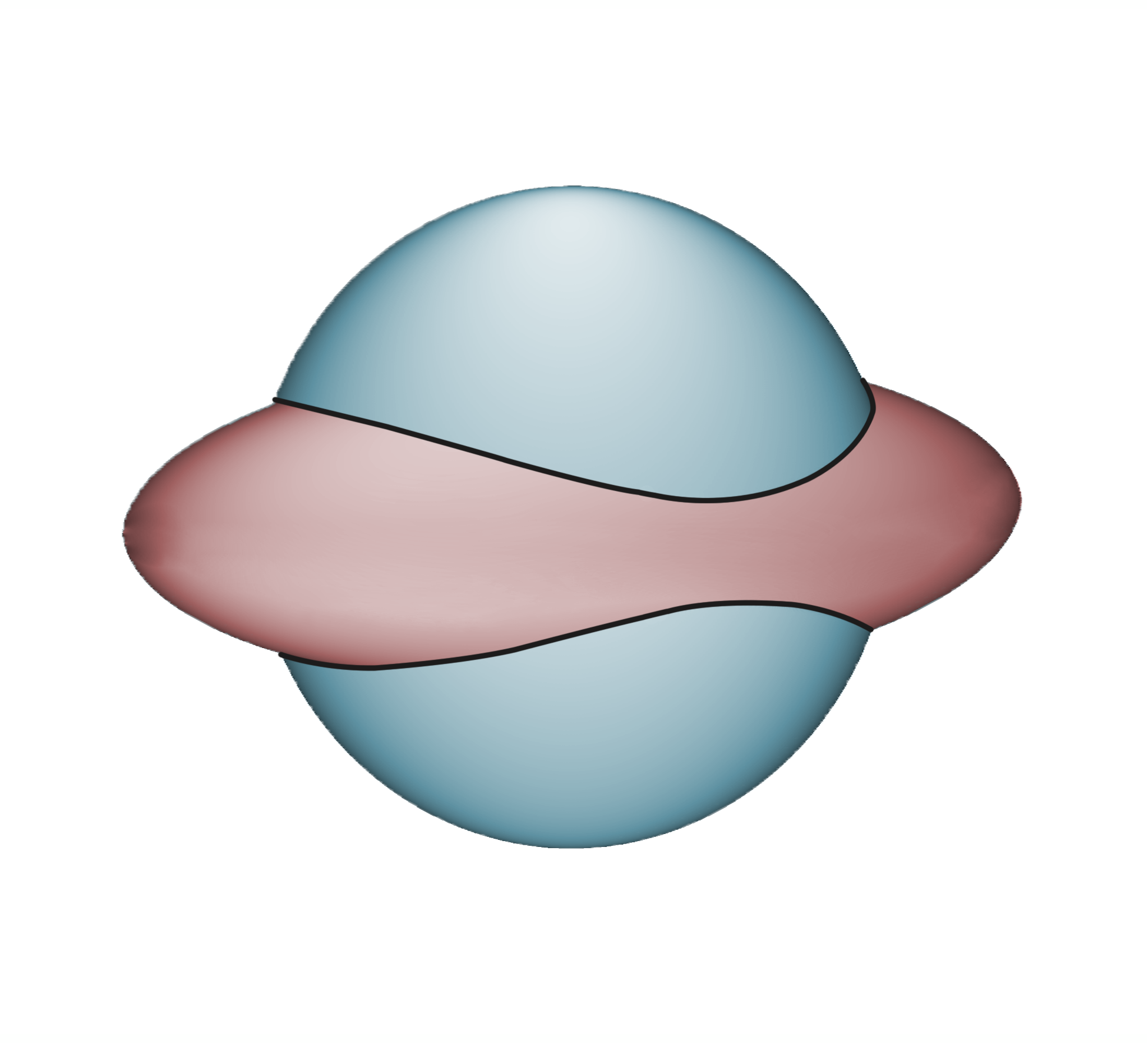}
     \caption{Three-dimensional illustration of the intersection between an ellipse and a sphere, studied here in $N$ dimensions. As the second semi-axis exceeds the sphere radius, the intersection transitions from two disjoint lobes to a single connected set -- this is the transition to reversals studied here.}
     \label{fig:ill_int}
 \end{figure}

In the problem examined below, one can easily show that the grand canonical description fails.
We consider the TEE in a square $(x,y)\in[0,\pi]^2 \eqqcolon \mathcal{D}$ with free-slip boundary conditions. 
This allows one to write the stream function as a double-sine series with real coefficients $\psi_{n,m}$
\begin{equation}
    \psi(x,y) = \sum_{m,n} \psi_{n,m} \sin(mx) \sin(ny), \label{eq:sin2expansion}
\end{equation}
with a truncation that retains $N$ modes $(m,n)$. \textcolor{black}{As described in section \ref{sec:cond} \ref{sec:cond_mic}, the summation in (\ref{eq:sin2expansion}) is over the set $K=\lbrace (m,n)\in \mathbb{N}_+^2| 0<\sqrt{m^2+n^2}<k_{max}  \rbrace$}. We again enumerate all retained modes by a single index $i$ as $(m(i),n(i))$, non-decreasing in $k_i\coloneqq \sqrt{m(i)^2+n(i)^2}$, i.e. $k_1=\sqrt{2}, k_2=k_3=\sqrt{5}, k_4=\sqrt{8},\dots$. 
We also define the more convenient variables $r_i=\psi_{m(i),n(i)} k_i/\sqrt{2}$, as in the previous section. Then energy and enstrophy conservation read, once again
\begin{equation}
    \sum\nolimits_{i=1}^N r_i^2 = \cE, \hspace{2cm} \sum\nolimits_{i=1}^N k_i^2 r_i^2 = \Omega.
    \end{equation}
For this system it is clear that if $\Omega < \cE k_2^2$, then the amplitude of the $r_1$ mode cannot be reduced to zero because that would correspond 
to a $\Omega\ge \cE k_2^2$ situation. Thus, if $r_1$ is positive/negative at $t=0$ it will remain positive/negative at all times \NEW{(note the importance of $\psi_{1,1}\in \mathbb{R}$ at this step in the argument)}. A 3-D geometric illustration of this result is shown in left most panel of figure \ref{fig:ill_int} where it is shown that the intersection of a sphere with an ellipsoid results in two disjoint lobes. This is in contradiction with the grand canonical description, which assumes a Gaussian PDF $\cP(\bu) \propto \exp\left[\sum (\alpha+\beta k_i^2) r_i^2 \right]$ and thus $r_i=0$ is always the most probable value for $r_i$.
It is, however, not an issue in the microcanonical ensemble, which follows the geometrical description illustrated in figure \ref{fig:ill_int}. For $\Omega\ge \cE k_2^2$, the amplitude $r_1$ can change sign (i.e. the large-scale flow can reverse) with a probability that becomes smaller and smaller as $\Omega$ approaches the critical value $\Omega_c=\cE k_2^2$ from above. Based on this insight, we define $\varepsilon$ by 
\begin{equation} 
\Omega\eqqcolon k_2^2 \cE (1+\varepsilon).
\end{equation}
At $\varepsilon=0$, the system undergoes a transition where reversals appear. 
We emphasize that $\varepsilon$ is different from $\epsilon$ used in the previous section. 
In particular, $\varepsilon$ may take both signs and need not be small. \textcolor{black}{The possibility of reversals in the large-scale circulation has been discussed previously in \cite{chavanis1996classification}.} In this section we explicitly calculate the reversal probability and its scaling with the deviation from onset $\varepsilon$, using the microcanonical description as before. 
 
Denote by $\mathbb{S}(\cE)$ the spherical shell in $N$ dimensions, with energy in $[\cE,\cE+d\cE]$ for infinitesimal $d\cE$. Similarly, denote by $\mathbb{E}(\Omega)$ the ellipsoidal shell in $N$ dimensions, with enstrophy in $[\Omega,\Omega+d\Omega]$ for infinitesimal $d\Omega$. We wish to compute the following microcanonical probability 
\begin{equation}
    P(r_1\in[z,z+dz]) = \frac{\textrm{Vol}\left(\left. r_i\in \mathbb{S}(\cE)\cap \mathbb{E}(\Omega) \right| r_1\in[z,z+dz] \right)}{\textrm{Vol}\left(\mathbb{S}(\cE)\cap \mathbb{E}(\Omega)\right)},
\end{equation}
or equivalently, the probability density $p(z)$, satisfying $P(r_1\in[z,z+dz])\eqqcolon p(z)dz$.\\
\textcolor{black}{Similar to section \ref{sec:cond}, we will denote $z=r_1$, $r_2=x\cos(\theta)$, $r_3=x\sin(\theta)$, $y^2=\sum_{i=4}^N r_i^2$. This gives}
\begin{align}
    z^2+x^2+y^2 =& \cE, \label{eq:energy_cons}\\
    k_1^2 z^2+k_2^2 x^2+q^2(\Phi)y^2 = &\Omega ,\label{eq:enstrophy_cons}
\end{align}
\textcolor{black}{where $q^2=q^2_{M+1}$ is given by eq. (\ref{eq:def_q}) with $M=3$, and the angles $\Phi$ are defined by adopting spherical coordinates for $(r_4,\dots,r_N)$, as described explicitly in the appendix \ref{sec:appB}.} 
 By eliminating $y$ and $x$ from (\ref{eq:energy_cons}) and (\ref{eq:enstrophy_cons}), respectively, one finds
 \begin{align}
     x^2 = \cE \underbrace{\frac{q^2(\Phi)-(1+\varepsilon)k_2^2}{q^2(\Phi)-k_2^2}}_{=a}  - \underbrace{\frac{q^2(\Phi)-k_1^2}{q^2(\Phi)-k_2^2}}_{=b} z^2  , \hspace{1cm}       y^2 = \frac{(k_2^2-k_1^2)z^2+\varepsilon k_2^2 \cE}{q^2(\Phi)-k_2^2}. \label{eq:y2_z2}
 \end{align}
 \textcolor{black}{These relations imply several important constraints, as described in detail in appendix \ref{sec:appB}. We highlight the following: for fixed $\varepsilon\geq 0$, there is a value $z_c(\varepsilon)$ of $|z|$, such that at $|z|\leq z_c(\varepsilon)$ all angles $\Phi$ are consistent with $x^2=\cE a-bz^2 >0$ in (\ref{eq:y2_z2}). In this case integrals over $\Phi$, which arise when computing $p(z)$, must be performed over the whole $(N-4)$-sphere. For $|z|>z_c$, only a non-trivial subset of the $(N-4)$-sphere satisfies $\cE a -bz^2>0$, which complicates $\Phi$ integration.}

 We proceed by fixing $z,\theta,\Phi$ and considering the $x,y$ plane. The intersection between the spherical energy shell and the ellipsoidal enstrophy shell in this plane is a parallelogram  of height 
 \begin{equation}
     \delta R = \frac{d \cE}{2\sqrt{\cE-z^2}}
 \end{equation} and base length $\delta L = \sqrt{(x_A-x_B)^2+(y_A-y_B)^2}$, where $A$ is a point at energy $\cE$ and enstrophy $\Omega$, while $B$ is a point at energy $\cE$ and enstrophy $\Omega+d\Omega$, as in the previous chapter. This situation is the one depicted in figure \ref{fig:circles_ellipses}. We take $d \cE, d \Omega$ infinitesimally small. The parallelogram area is
 \begin{equation}
     \delta A(z,\boldsymbol{\varphi}) = \delta R \delta L.
 \end{equation}
 The base length $\delta L$ satisfies 
 \begin{align*}
     \delta L^2 =&  (x_A-x_B)^2+(y_A-y_B)^2 \\
                =&  \frac{(x_A^2-x_B^2)^2}{4x^2}+\frac{(y_A^2-y_B^2)^2}{4y^2}  \\
                \stackrel{(\ref{eq:y2_z2})}{=}&  \frac{d \Omega^2}{4(q^2-k_2^2)^2} \frac{\cE-z^2}{x^2y^2}.
 \end{align*}
 Putting these expressions together, we can compute the sought-after probability density
 \begin{equation}
     p(z) \propto \int \delta A x  d\theta y^{N-4} d\Phi = \frac{d \cE d\Omega}{4} 2\pi \int \frac{y^{N-5}}{(q^2-k_2^2)} d{\Phi},
 \end{equation}
 where the normalisation is omitted. Using (\ref{eq:y2_z2}) to express $x,y$ as a function of $z$ and $\Phi$ gives
 \begin{equation}
     p(z) \propto \left((k_2^2-k_1^2)z^2+\varepsilon k_2^2 \cE\right)^{\frac{N-5}{2}} \underbrace{\int_{S(z,\varepsilon)} \left( q^2(\Phi)-k_2^2\right)^{\frac{3-N}{2}}}_{\eqqcolon f(z,\varepsilon)} d\Phi 
     \label{eq:pdf_less_simple}
 \end{equation}
 where $S(z,\varepsilon)$ denotes the subset of the $N-4$-dimensional ${\Phi}$ unit sphere contributing to the integral at a given $z$ and $\varepsilon$. 
First consider $0 \leq \varepsilon <\varepsilon_{c}$ and $|z|\leq |z_{c}(\varepsilon)|$.  In this case, $S(x,\varepsilon)=S_{N-4}$ is the whole unit $N-4$-sphere, and the $\Phi$ integral gives a $z$-independent constant. Thus, we obtain
 \begin{equation}
     {p(z)\propto \left(\sqrt{(k_2^2-k_1^2)z^2+\varepsilon k_2^2 \cE}\right)^{N-5} } \label{eq:pdf_simple}
 \end{equation}
The result does not include normalisation, which will depend on $\cE$, $\varepsilon$ and the $k_i$. 
\textcolor{black}{Eq. \ref{eq:pdf_simple} was verified by a Monte-Carlo computation, uniformly sampling from the spherical shell $\mathbb{S}(\cE)$, retaining only the points in the intersection with $\mathbb{E}(\Omega)$ (not shown).}  For small $\varepsilon>0$, it implies that 
\begin{equation}
    p(z=0)\propto \varepsilon^{\frac{N-5}{2}} \label{eq:p0_scaling}
\end{equation} 
(at small $\varepsilon$, the normalisation becomes independent of $\varepsilon$ to leading order). The bottleneck \textcolor{black}{(the term is used here without any relation to the bottleneck phenomenon referenced in the introduction)} illustrated in figure \ref{fig:ill_int} thus becomes thinner as $\epsilon$ decreases and as $N$ increases. Moreover, for small $\varepsilon>0$, there is a power-law range $p(z)\propto |z|^{N-5}$ at intermediate $|z|$, which becomes steeper as $N$ increases. It thus becomes less likely to reach states close to $z=0$ as $N$ increases. In the above calculation, the two real modes $(1,2)$ and $(2,1)$ are associated with the second wavenumber $k_2$. If instead, there are $\tilde{M}$ degrees of freedom associated with $k_2$ (e.g. $\tilde{M}=1$ in a non-square rectangular free-slip domain), then one can show that (\ref{eq:p0_scaling}) is replaced by $p(z=0)\propto \varepsilon^{\frac{N-\tilde{M}-3}{2}}$, reproducing (\ref{eq:p0_scaling}) for $\tilde{M}=2$. We further note that eq. (\ref{eq:pdf_simple}) also applies to TEE flow in a channel with mixed free-slip-periodic boundary conditions as studied in \cite{dallas2020transitions}, with $k_1=1$, $k_2=\sqrt{2}$.
\\
 
\noindent  If either (i) $\varepsilon< \varepsilon_{c}$, $|z|>|z_{c}|(\varepsilon)$, or (ii) $\varepsilon   \geq \varepsilon_{c}$, then the integration boundaries are $z$-dependent and $p(z)$ in (\ref{eq:pdf_simple}) is modified by a non-trivial $z$-dependent factor $f(z,\varepsilon)$ given in eq. (\ref{eq:pdf_less_simple}). \textcolor{black}{The integral can in principle be computed numerically for small $N$, which we have verified for the for the simplest non-trivial case $N=5$ (not shown)}. However, this becomes increasingly costly for higher values of $N$.
If $\varepsilon>\varepsilon_c$, then $f(z,\varepsilon)$ decreases strictly monotonically as $|z|$ increases, competing against the square root term, which increases from $z=0$. For sufficiently large $\varepsilon$, the PDF develops a maximum at $z=0$. Eventually, $p(z)$ approaches a Gaussian centered on $z=0$, as is seen in \cite{shukla2016statistical}. Only in that special case may one attempt to describe the reversal statistics using the canonical ensemble, while the present microcanonical description also captures the behaviour of the system close to $\varepsilon=0$. 

\subsection{Comparison with numerical simulations}
\begin{figure}
    \centering
    \includegraphics[width=0.45\textwidth]{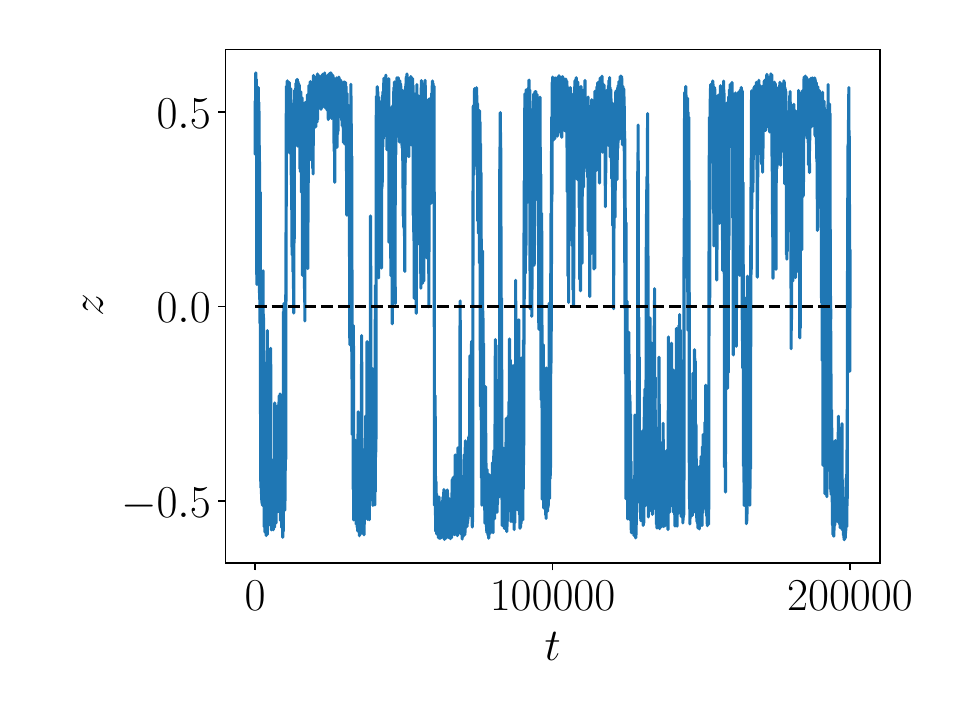}
    \includegraphics[width=0.45\textwidth]{pdfs_and_fits.pdf}
    \caption{Left: time series of the amplitude $z$ of the large-scale mode. Right: PDF $p(z)$ versus $z$ for $\epsilon=0.3,0.23,0.14,0.08,0.05$ (top to bottom at $z=0$). The black dashed line indicates the theoretically predicted functional form, the prefactor is determined by fitting. The endpoints of the dashed lines are given by $z=\pm z_c(\varepsilon)$ defined in eq. (\ref{eq:xc_of_epsilon}). Beyond this point, eq. (\ref{eq:pdf_simple}) ceases to be valid, and is replaced by (\ref{eq:pdf_less_simple}) which is harder to evaluate. 
    }
    \label{fig:pdf_reversals}
\end{figure}

\begin{figure}[h]
    \centering
    \includegraphics[width=0.45\textwidth]{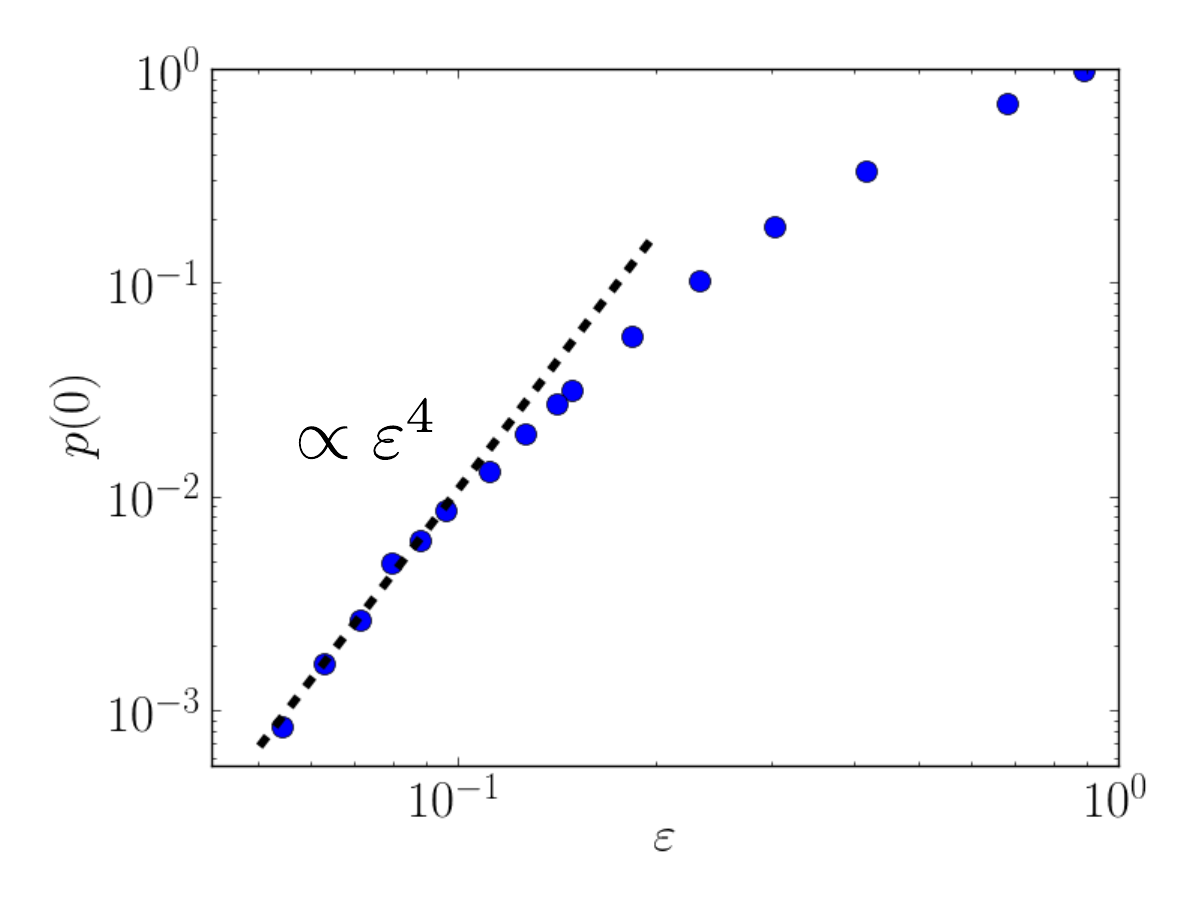}
    \includegraphics[width=0.45\textwidth]{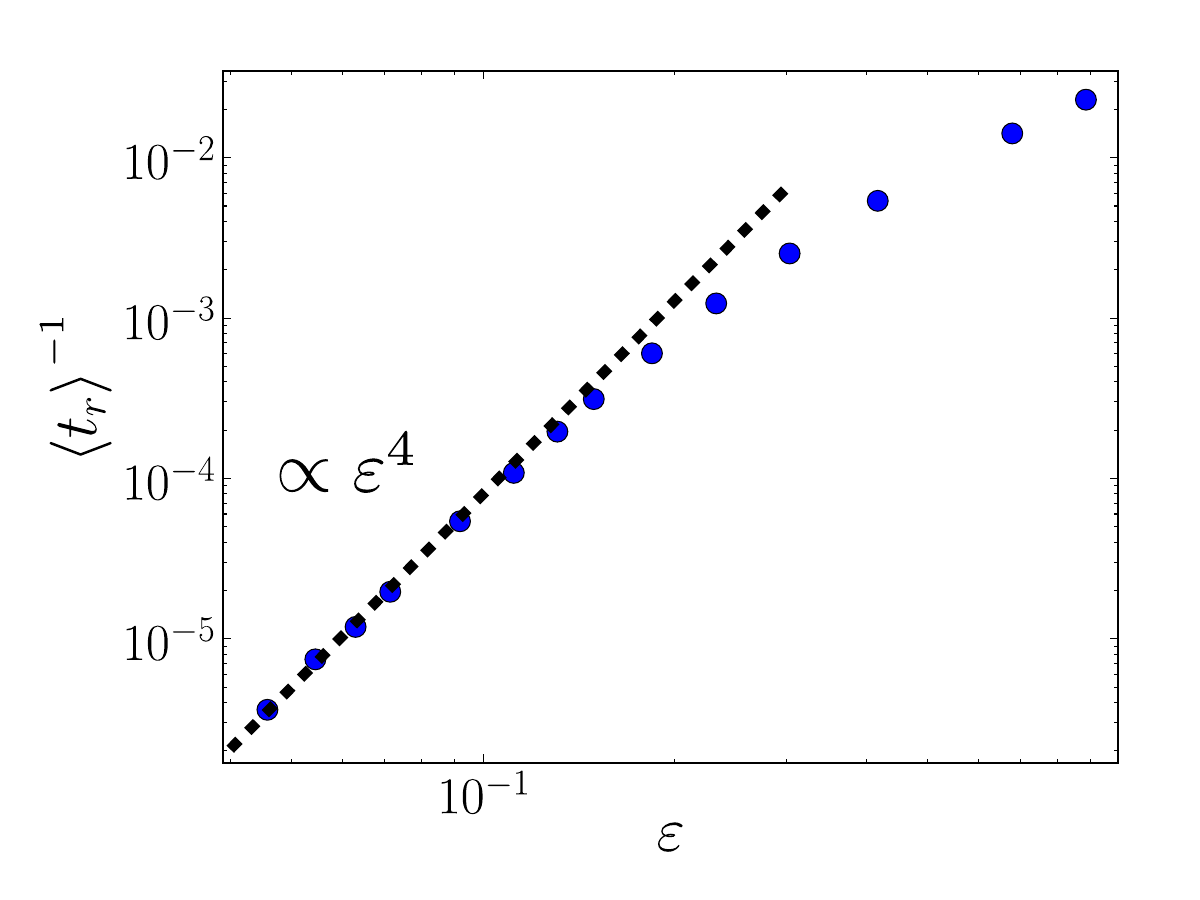}
    \caption{
    Left: $p(z=0)$ versus $\varepsilon$. The dashed line indicates the scaling $\varepsilon^4$ predicted theoretically.
    Right: Inverse of mean reversal time as a function of $\varepsilon$. The scaling at small $\varepsilon$ is proportional to $p(0)$.}
    \label{fig:passage_times}
\end{figure}

In this section, we confront the analytical predictions derived above with numerical solutions of the minimal 13-mode model that is given explicitly in \cite{shukla2016statistical}. This minimal model corresponds to the TEE in the square domain with free-slip boundaries and $k_{max}=2\sqrt{5}$. We initialise simulations in a state with $E(k)=\frac{1}{2(\alpha+\beta k^2)}$. For fixed $\beta$, we vary $\alpha$, and in each case normalise such that the total energy is $\cE=1/2$. Thus we generate states with equal $\cE$, but different $\Omega$, or equivalently, different $\varepsilon$. We use a fourth-order Runge-Kutta scheme to integrate the 13-mode TEE for long times (up to $O(10^{11})$ time steps). From this, we obtain time series such as the one shown in the left panel of figure \ref{fig:pdf_reversals}, from which we may construct histograms of $z$. The right panel of figure \ref{fig:pdf_reversals} shows the resulting PDF, $p(z)$, for different values of $\varepsilon$. One observes that the value $p(z=0)$ decreases with $\varepsilon$ and the weight of the PDF shifts to larger $|z|$. An excellent agreement is found between the theoretical predictions, shown by the black dashed lines in figure \ref{fig:pdf_reversals}, and the results of the numerical integration. The normalisation constant, which is the unique parameter not predicted by the theory, was determined by fitting the theoretical prediction to the data. We reiterate that Kraichnan's canonical description (\ref{eq:kraichnan_canonical_pdf}) is inadequate here, given the bi-modal shape that is far from being Gaussian, which indicates that the microcanonical description is required. While the exact normalisation constant is not given by (\ref{eq:pdf_simple}), the scaling prediction of equation (\ref{eq:p0_scaling}) for $N=13$ is that $p(z=0)\propto \epsilon ^4$. This is confirmed in the left panel of figure \ref{fig:passage_times}. Geometrically, the fraction of the intersection volume close to $z=0$ shrinks rapidly as $\epsilon\to 0 ^+$. This suggests that transitions from one lobe to the other will be controlled by the bottleneck illustrated in figure \ref{fig:ill_int}. 

A further characteristic of the system is the average waiting time $\langle t_r\rangle $ between two reversals, shown in the right panel of figure \ref{fig:passage_times} for different $\varepsilon$. It is found that the inverse of $\langle t_r\rangle $ scales as $\varepsilon^{\frac{N-5}{2}}$ at small $\varepsilon$, just as $p(0)$. This translates the aforementioned fact that the width of the bottleneck controls the reversals for small $\varepsilon$. \textcolor{black}{It is interesting to note that in the Kramers problem \cite{kramers1940brownian}, which consists of the escape of a Brownian particle from a potential well, a similar relation is found between the mean first-escape time and the probability to be at the edge of the well. There, the constant of proportionality can be computed in terms of the curvature of the potential. In the TEE, by contrast, there is no underlying potential which generates the dynamics. Therefore, a priori, the constant of proportionality cannot be derived in the same way as for the Kramers problem. The chaotic motion of the higher-dimensional Truncated-Euler dynamics may possibly be modeled by noise. Alternatively, instanton theory, which has already proven to be a powerful tool for studying transitions in multistable hydrodynamic systems \cite{bouchet2014langevin,laurie2015computation}, provides a promising approach to studying the reversal statistics here.} 

Since $p(0)\propto \varepsilon^{(N-5)/2}$ for small $\varepsilon$, the PDF $p(z)$ near $z=0$ will converge to zero as $N\to \infty$ \textcolor{black}{with $\varepsilon$ fixed}. This points to a question of non-commuting limits. Taking $\varepsilon\to 0^+$ first and then increasing $N$ will likely yield a different result. Specifically, first taking $N$ to be large and then increasing $\varepsilon$ from $\varepsilon =0$ must be expected to display an onset of reversals at $\varepsilon=\varepsilon_{thr}>0$. By contrast, for finite $N$ the onset is at $\varepsilon=0$, even though, as described by \cite{shukla2016statistical}, an ergodicity delay takes place for small $\varepsilon$, which becomes more severe with increasing $N$ as our calculation indicates. 

The strict absence of reversals at $\varepsilon<0$ is related to the exact conservation properties of the TEE. A forced-dissipative Navier-Stokes flow with the same average energy and enstrophy values close to $\varepsilon=0$ may behave differently, since the conservation laws do not apply exactly, and energy and enstrophy always fluctuate. Nonetheless, our result on average reversal times may potentially have some relevance for experiments \cite{herault2015experimental,gallet2012reversals}, since it allows one to relate the number of modes in the system to an experimentally simple-to-measure quantity. In an experiment, if one controls the average energy and enstrophy of the flow, then one may hope to infer information on the effective number of modes active in the system by measuring reversal time statistics. In a realistic turbulent flow, the truncation is related to viscosity. 

\section{Conclusions}
We have provided two examples of explicit microcanonical computations, involving the exact solution of phase space volume integrals
\NEW{for the TEE system}. In the case of a strongly condensed TEE flow, we showed that the microcanonical average energy spectrum is identical to Kraichnan's canonical prediction at leading order, for any number of modes. In the second example, we extended the results of \cite{shukla2016statistical} and explicitly computed the functional form of the PDF of the large-scale mode $z$ of TEE flow confined in a square domain with free-slip boundaries. The prediction for the PDF in confined TEE flow was validated using a minimal 13-mode model. Our theoretical results on free-slip flow in a square domain also apply to the mixed free-slip-periodic flow studied in \cite{dallas2020transitions}. We further analysed the statistics of waiting time between reversals. In particular we observe numerically that the inverse of the mean time between reversals scales as $\varepsilon^{\frac{N-5}{2}}$ with the distance from threshold $\varepsilon$. This is proportional to the scaling of $p(z=0)$, depending strongly on $\varepsilon$ and the number of modes $N$. While our TEE-based computation does not take into account forcing and dissipation, it was established by \cite{shukla2016statistical,dallas2020transitions} that many properties of the large scales of Navier-Stokes flow in the same domain are well described by the TEE equations. At a practical level, reversal times are easily accessible in experiments such as \cite{herault2015experimental,gallet2012reversals}. Thus, a potential link with experiments could be made by measuring transition times for flows with different energy and enstrophy, which may be controlled by the forcing in an experiment. Our result relates the number of modes $N$ to the mean reversal time. Thus, one may be able to deduce an effective number of active dynamical modes in a laboratory flow from simply measuring reversal times.

This study provides a first step towards an explicit, geometric microcanonical theory of TEE flows, thus complementing the impressive existing body of literature on the statistical mechanics of turbulent flows. Future studies should aim to extend the microcanonical results presented here to 3-D TEE flows conserving both energy and helicity. Aperiodically reversing flows are observed in realistic geophysical flows, for instance in the form of the Quasi-Biennial Oscillation \cite{baldwin2001quasi} \NEW{and reversals of the large-scale magnetic field in dynamo flows \cite{berhanu2007magnetic}}. Similar transitions of the large-scale dynamics occur in many models for the dynamics of \NEW{geo- and astrophysical} flows, such as the primitive and quasi-geostrophic equations. If aspects of these transitions can be described using a TEE-type reduction, then a microcanonical approach similar to the one described here could be a useful tool for understanding statistical properties of these transitions. \textcolor{black}{An interesting possible extension of the work presented here, which goes beyond the dichotomy of choosing between the microcanonical and canonical ensembles, is to consider generalised canonical ensembles such as the ones formulated in \cite{costeniuc2006generalized}. }


\dataccess{This article has no additional data.}
\aucontribute{All authors participated in the analytical computations. AvK drafted the manuscript and performed the numerical simulations. All authors read, edited and approved the manuscript.}
\competing{The authors declare that they have no competing interests.}
\funding{This  work  was  supported  by  Agence  nationale  de  la  recherche  (ANR  DYSTURB  project  No. ANR-17-CE30-0004).}
\ack{\textcolor{black}{We would like to thank three anonymous referees for their detailed comments, which have helped to improve the clarity of this work significantly.} This work was granted access to HPC resources of MesoPSL financed by Region Ile de France and the project Equip@Meso (reference ANR-10-EQPX-29-01) of the  programme  Investissements  d'Avenir  supervised  by  Agence  Nationale  pour  la  Recherche  and  HPC resources of GENCI-TGCC \& GENCI-CINES (Projects No.  A0070506421,  A0080511423,  A0090506421). AvK  was supported  by Studienstiftung des deutschen Volkes. We would like to also thank E. Stavrakopoulou for her help with the figure graphics and Nikolas Claussen for pointing out the spinning top analogy. }


\begin{appendix}
\section{Details on the condensate case}
\label{sec:appA}
\subsection{Coordinate transform}
In order to be able to perform integrals efficiently, we define two sets of spherical coordinates in terms of angle variables  $\Theta=(\theta_1,\theta_2,\dots,\theta_{_{M-1}})$ and $\Phi=(\phi_{_{M+1}},\phi_{_{M+2}},\dots,\phi_{_{N-1}})$. If $M=1$, we simply let $r_1=x$ and no angles $\Theta$ are required. If $M>1$, \NEW{then the first $M$ variables are}
\begin{eqnarray}
r_n =& x \left(\prod_{i=1}^{n-1}\sin(\theta_i) \right) \cos(\theta_n) =& x g_{n}(\Theta) \hspace{1cm}\text{for }1\leq n\leq M-1,\\
r_{_M} =& x  \left(\prod_{i=1}^{M-1}\sin(\theta_i) \right)\qquad \hspace{0.25cm} =& x g_{_{M  }}(\Theta).
\end{eqnarray}
\NEW{The remaining $N-M$ variables are given by} 
\begin{eqnarray}
r_n =& y \left(\prod_{i=M+1}^{n-1} \sin(\phi_i) \right)\cos(\phi_n) =& y f_n(\Phi) \hspace{0.75cm} \text{for } M+1\leq n\leq N-1,\\
r_{_N} =& y \left(\prod_{i=M+1}^{N-1} \sin(\phi_i) \right) \qquad \hspace{0.25cm}=& y f_{_N}(\Phi).
\end{eqnarray}
The volume element $dV$ is given by
\[  dV = x^{_{M-1}} y^{_{N-M-1}} \, dx \, dy \, d\Theta  \, d\Phi_{M+1}   \]
with
\begin{eqnarray}
d\Theta &=& \sin^{_{M-2}}(\theta_1)\sin^{_{M-3}}(\theta_2) \dots \sin(\theta_{_{M-2}}) d\theta_1\, d\theta_2\dots d\theta_{_{M-1}}, \\
d\Phi_{_{M+1}}   &=& \sin^{_{N-M-2}}(\phi_{_{M+1}})\sin^{_{N-M-3}}(\phi_{_{M+2}}) \dots \sin(\phi_{_{N-2}}) d\phi_{_{M+1}}\,   d\phi_{_{M+2}} \dots d\phi_{_{N-1}},\qquad
\end{eqnarray}
(understanding that $d\Theta\coloneqq 1$ for $M=1$, \textcolor{black}{and letting $N\geq M+2$, so that $\Phi$ contains at least one angle variable}), see \url{https://en.wikipedia.org/wiki/N-sphere#Spherical_coordinates}\textcolor{black}{, or alternatively} \cite{blumenson1960derivation}. 
\textcolor{black}{We stress that the angles $\Theta$ drop out of equation (\ref{eq:sphell}), while a non-trivial dependence on the angles $\Phi$ remains. As a consequence, $\Theta$ integrations will be trivial, while those over $\Phi$ must be performed iteratively. Therefore, we define the differential $d\Theta$ without an index, but $d\Phi_{M+1}$ with an index to keep track of the iterations.}
In terms of the spherical coordinates given above, the quantity $q_{M+1}(\Phi)$ defined in equation (\ref{eq:def_q}) satisfies
\begin{equation}
    q^2(\Phi) = \sum_{n=M+1}^{N-1} k_n^2 \left(\prod_{i=M+1}^{n-1} \sin^2(\phi_i)\right) \cos^2(\phi_n)  + k_N^2\prod_{i=M+1}^{N-1} \sin^2(\phi_i).  
\end{equation}
\subsection{Angular integration}
With the infinitesimal area element derived in the main text, we can now perform the angular integration. We first consider the N-dimensional volume of the intersection $\mathcal{Z}$, given in (\ref{eq:z}) as
\beq 
\mathcal{Z} = \int \frac{d \cE d \Omega}{4 \epsilon \cE k_1 \sqrt{q_{_{M+1}}^2-k_1^2} }. x^{M-1}y^{N-M-1} d\Phi_{_{M+1}} d\Theta.
\eeq 
Note that the integrand is independent of $\Theta$. After substituting the expressions for $y,x$ from (\ref{eq:xyq}) and integrating over the angles $\Theta$, the integral becomes
\begin{eqnarray}
\mathcal{Z} = \frac{1}{4}S_{M-1} (\epsilon k_1)^{N-M-2} \cE^{\frac{N}{2}-2} d \cE d \Omega
               \underbrace{ \int \left(\frac{1}{q_{_{M+1}}^2-k_1^2}\right)^{\frac{N-M}{2}}d\Phi_{_{M+1}}}_{\eqqcolon I} \label{eq:deltaV} 
\end{eqnarray}
where $S_{M-1}$ is the surface of the unit-radius $(M-1)$-sphere ($S_0\coloneqq 1$). 
Integrating over $\phi_{_{M+1}}$, making the substitution $u = \sqrt{\frac{q_{M+2}^2-k_{M+1}^2}{k_{M+1}^2-k_1^2}} \tan(\phi_{M+1})$, gives 
\beq 
I = \frac{\left( \int \left(\frac{1}{q_{_{M+2}}^2-k_1^2}\right)^{\frac{N-M-1}{2}}
      d\Phi_{_{M+2}} \right)}{(k_{_{M+1}}^2-k_1^2)^{1/2}}  \left( \int \left(\frac{1}{ 1  + u^2}\right)^{\frac{N-M}{2}}
  u^{_{N-M-2}}  du \right).  \label{eq:intI}
\eeq 
As shown below, further simplifications are not necessary for obtaining the final result.

The integrals (\ref{eq:r2}) can be performed by a procedure similar to that just presented for eq. (\ref{eq:z}).
Here two cases must be distinguished. For $i=1,\dots,M$, to leading order, we need to compute
\begin{eqnarray}
\langle r_i^2 \rangle = \frac{1}{\mathcal{ Z}} \int  \frac{g_i^2(\Theta) d \cE d \Omega}{4 \epsilon  \cE k_1 \sqrt{q^2_{M+1}-k_1^2}} x^{M+1} y^{N-M-1} d\Phi_{M+1} \label{eq:In_i_lt_M} d\Theta. \end{eqnarray}
For $i=M+1,\dots,N$, the integral to be computed is given, to leading order, by
\begin{eqnarray}
\langle r_i^2 \rangle = \frac{1}{\mathcal{ Z}} \int  \frac{f_i^2(\Phi_{_{M+1}}) d \cE d \Omega}{4 \epsilon  \cE k_1 \sqrt{q^2_{M+1}-k_1^2}} x^{M-1} y^{N-M+1} d\Phi_{M+1}.
\end{eqnarray}
We first explicitly consider $i=M+1$.
\begin{eqnarray}
\langle r_{_{M+1}}^2 \rangle &=& \frac{1}{\mathcal{Z}} \int r_{_{M+1}}^2 dV \\
&=& \frac{1}{\mathcal{Z}} \int \frac{d \cE d \Omega}{4 \epsilon \cE k_1 \sqrt{q_{_{M+1}}^2-k_1^2} } x^{M-1}y^{N-M+1} \cos^2(\phi_{_{M+1}}) d\Phi_{_{M+1}} d\Theta \notag \\
&=& \frac{1}{4 \mathcal{Z}}S_M (\epsilon k_1)^{N-M} \cE^{\frac{N}{2}-1} d \cE d \Omega \underbrace{
                \int \left(\frac{1}{q_{_{M+1}}^2-k_1^2}\right)^{\frac{N-M}{2}+1} \cos^2(\phi_{_{M+1}}) d\Phi_{_{M+1}}}_{\eqqcolon J}\qquad  \label{eq:I_Mplus1}
\end{eqnarray}
The last integral $J$ can again be calculated by use of the substitution $u=\sqrt{\frac{q_{M+2}^2-k_1^2}{k_{M+1}^2-k_1^2}} \tan(\phi_{M+1})$,
\beq 
J= \frac{1}{(k_{_{M+1}}^2-k_1^2)^{3/2}} \left( \int \left(\frac{1}{q_{_{M+2}}^2-k_1^2}\right)^{\frac{N-M-1}{2}}
      d\Phi_{_{M+2}} \right)   \left( \int \left(\frac{1}{ 1  + u^2}\right)^{\frac{N-M}{2}+1}
  u^{_{N-M-2}}  du \right) \hspace{0.775cm} \label{eq:intJ}.
\eeq 
So, combining eqns. (\ref{eq:deltaV}), (\ref{eq:intI}) and (\ref{eq:I_Mplus1}), (\ref{eq:intJ}),  we finally have
\beq
\langle r_{_{M+1}}^2 \rangle = \frac{ \cE(\epsilon k_1)^2}{k_{M+1}^2-k_1^2} \,\,
\left(\frac{ \int \left(\frac{1}{ 1  + u^2}\right)^{\frac{N-M}{2}+1} u^{_{N-M-2}}  du }{\int \left(\frac{1}{ 1  + u^2}\right)^{\frac{N-M}{2}} u^{_{N-M-2}}  du}\right) = \frac{\cE(\epsilon k_1)^2}{k_{M+1}^2-k_1^2} \underbrace{\frac{\Gamma\left(\frac{N-M}{2}\right)}{2\Gamma\left(\frac{N-M}{2}+1\right)}}_{=(N-M)^{-1}} \label{eq:final_res_Mp1}
\eeq

\noindent
To find $\langle r_i^2 \rangle $ for $i=M+2,\dots,N$, we may simply choose a different set of spherical coordinates with $k_{M+1}\to k_i$ at the outset. This amounts to replacing $k_{M+1}$ by $k_i$ in (\ref{eq:final_res_Mp1}). Hence, for all $i > M$
\beq
E(\bk_i)=
\langle r_i^2 \rangle =  \frac{\epsilon^2 \cE k_1^2 }{(N-M)(k_i^2-k_1^2)} 
\label{eq:In_i_gt_M}
\eeq
For $i=1,\dots, M$, all values of $i$ give the same result by symmetry (all $k_i$ being equal for $i\leq M$). Conservation of energy thus yields, at leading order,
\beq
E(\bk_i)= \langle {r_i}^2 \rangle =  \frac{1}{M}\left( \cE - \sum_{j=M+1}^{N} \langle r_j^2 \rangle \right)  = \cE/M + O(\epsilon^2). \label{eq:In_result_lt_M}
\eeq

\section{Details on the free-slip case}
\label{sec:appB}
\subsection{Coordinate transformation}
In order to simplify the integration, we transform from the $r_i$ to the following set of coordinates,
\begin{align*}
    r_1 =& z, \\
    r_2 =& x \cos(\theta) &&\\ 
    r_{3} =& x \sin(\theta)&&\\ 
    r_n =& y \left(\prod\nolimits_{i=4}^{n-1} \sin(\phi_i) \right) \cos(\phi_n) =& y f_n({\Phi})& \hspace{1cm}\text{for } 4\leq n\leq N-1, \\
    r_N =& y \left(\prod\nolimits_{i=4}^{N-1} \sin(\phi_i) \right) \qquad \hspace{0.45cm} =& y f_N({\Phi})&,
\end{align*}

with \textcolor{black}{ ${\Phi} = (\phi_{4},\dots,\phi_{N-1})$. \textcolor{black}{We restrict our attention to $N\geq 5$, so that $\Phi$ always contains at least one angle variable.} The variable names are chosen by analogy with section \ref{sec:cond}. }

\subsection{Constraints}
The expressions for $x$ and $y$ given in eq. (\ref{eq:y2_z2}) imply several important constraints.
 \begin{enumerate}
     \item For $\varepsilon<0$, imposing $y^2\geq0$ gives 
     \begin{equation}
         z^2\geq z_{min}^2 = |\varepsilon|k_2^2  \cE/(k_2^2-k_1^2).
     \end{equation}
     This is consistent with the geometrical insight. It implies $p(z=0)=0$ for $\varepsilon\leq 0$. A transition from no reversals to reversals occurs at $\varepsilon=0$.
     \item For $\varepsilon\geq 0$, $a\leq1$ and $b> 1$ in (\ref{eq:y2_z2}). Further, $a$ and $-b$ increase as $q^2$ increases. This implies that in order for $x^2=a \cE -b z^2$ to be greater than or equal to zero for all $\Phi$, one must have
     \begin{equation}z^2 \leq z_{c}(\varepsilon)^2\coloneqq \cE  \left.\left(k_{4}^2-k_2^2(1+\varepsilon)\right)\right/\left(k_{4}^2-k_1^2\right).\label{eq:xc_of_epsilon}\end{equation}
     As long as this is satisfied, integrals over the angles $\Phi$, which need to be performed for computing $p(z)$, are over the whole unit $(N-4)$-sphere.
     \item In order for $z_{c}^2\geq0$, it is necessary that
     \begin{equation}
         \varepsilon\leq \varepsilon_{c}= k_4^2/k_2^2-1.
     \end{equation}
     If $\varepsilon>\varepsilon_c$ or $|z|>|z_c|$, the $\Phi$ integration is nontrivial due to $z$-dependent integration limits.
     \item For a given $\varepsilon$, there is a value $z_{max}^2$ of $z^2$ such that $x^2\geq 0$ in (\ref{eq:y2_z2}) cannot be satisfied for any $\Phi$ for $|z|>z_{max}$. The PDF $p(z)$ vanishes for $z\geq z_{max}$. It is given by
     \begin{equation}
     z_{max}^2 = \cE  \left.\left(k_N^2-(1+\varepsilon)k_2^2\right)\right/ \left(k_N^2-k_1^2\right).
     \end{equation}
\end{enumerate}
\end{appendix}

\vskip 6pt

\enlargethispage{20pt}

\bibliography{biblio} 
\bibliographystyle{ieeetr}

\end{document}